\documentclass[structabstract]{aa}
 \usepackage{graphicx}
 \usepackage{txfonts}

\begin{document}

   \title{The follow-up EVN observations of twelve GPS radio sources\\ at 5 GHz}

   \author{L. Cui
           \inst{1,2},
           X. Liu
           \inst{1}\fnmsep\thanks{Corresponding author: e-mail: liux@uao.ac.cn},
           J. Liu
           \inst{1,2},
            H. -G. Song
           \inst{1,2},
           \and
           Z. Ding\inst{1,2}
          }
    \institute{Urumqi Observatory, the National Astronomical Observatories, CAS,
               40-5 South Beijing Road, Urumqi 830011, P.R. China
             \and Graduate University of the Chinese Academy of Sciences, Beijing 100049, P.R. China
              }
    \date{Received / Accepted }


  \abstract
   {}
   {Gigahertz peaked spectrum (GPS) radio
   sources are compact extragalactic radio sources, assumed to be young radio-loud active galactic nuclei
   and ideal objects for studying the early evolution of
   extragalactic radio sources. The Very Long Baseline Interferometry (VLBI)
   observation with high resolution is vital for studying the structure
   of these compact sources. }
   {We defined a sub-sample of twelve GPS sources which have not been
   observed with the VLBI before, from the Parkes half-Jansky
   sample, and carried
out VLBI observations at 1.6 GHz and 5 GHz with the European VLBI
Network (EVN) in 2006 and 2008, respectively, to classify the
source structure and
   to find compact symmetric objects (CSOs). Additionally, we carried out the 4.85
   GHz flux density observations for these sources with the Urumqi 25-m
   telescope between the years 2007 and 2009 to study whether there is any variability
   in the total flux density of the GPS sources.}
  {The results of the 5 GHz VLBI observations and total flux densities of these sources are
  presented in this paper. From the VLBI morphologies, the spectral indices of components
  and the total flux variability of the twelve targets, we firmly classify three sources J0210+0419, J1135$-$0021, and J2058+0540 as CSOs, and
  classify J1057+0012, J1203+0414, and J1600$-$0037 as core-jet sources. The others J0323+0534, J0433$-$0229, J0913+1454,
  J1109+1043, and J1352+0232 are labelled CSO candidates, and
  J1352+1107 is a complex feature. Apart from core-jet sources, the
total flux densities of the CSOs and candidates are quite stable
at 5 GHz both during a long-term of $\sim$20 years relative to the
PKS90 data and in a period between 2007 and 2009. The total flux
densities are resolved-out by more than 20\% in the 5 GHz VLBI
images for 6 sources, probably because of diffuse emission. In
addition, we estimated the jet viewing angles ($\Theta$)
  for the confirmed CSOs by using the double-lobe flux ratio of the sources, the result being indicative
  of relatively large $\Theta$ for the CSOs.}
   {}


\titlerunning{EVN observations of twelve GPS radio sources at 5 GHz}
   \authorrunning{L. Cui, X. Liu et al.}
\maketitle

\section{Introduction}

Gigahertz peaked spectrum (GPS) radio sources are powerful
($P_{\rm1.4~GHz}\geq\rm10^{25}\,W\,Hz^{-1}$) and compact
extragalactic radio sources characterized by a convex radio
spectrum peaking at between 0.3 and a few GHz (see O'Dea 1998 for
a review). Sources that peak at $>$ 5 GHz are called high
frequency peakers (HFP, Dallacasa et al. 2000), which are
understood to be even more compact and younger. The spectrum
turnover of the GPS sources is caused by synchrotron
self-absorption (de Vries et al. 2009), in general, and it is also
possible that free-free absorption plays a role in producing the
spectral turnover in some cases (e.g., Luo et al. 2007; Orienti \&
Dallacasa 2008).

Extended emission in GPS source is uncommon especially for GPS
galaxies, most of which are truly compact ($< 1\,kpc\,h^{-1}$)
(Stanghellini et al. 2005). The compactness of GPS sources is most
likely due to their youth ($< 10^4$ years) according to a spectral
ageing analysis (Murgia et al. 2003). The youth scenario is
supported by the hotspot kinematics in a couple of GPS sources
that are also identified as compact symmetric objects (CSOs)
(Giroletti \& Polatidis 2009). A unification model assumes that
GPSs/CSOs will evolve into compact steep spectrum (CSS) sources or
medium-sized symmetric objects (MSOs, $1-20\,kpc\,h^{-1}$), which
in turn, evolve into large radio galaxies ($> 20\,kpc\,h^{-1}$),
i.e., FRI/II radio sources (Fanti et al. 1995, 2009; Snellen et
al. 2000).

Furthermore, the GPS sources are associated with either galaxies
or quasars. The GPS quasars usually exhibit some form of core-jet
structure, while the GPS galaxies often show CSO-like or compact
doubles (CD)(e.g., Orienti et al. 2006, Liu et al. 2007),
suggesting that the GPS galaxies have relatively large jet-viewing
angles. The GPS galaxies have very low polarization (less than
0.5\% at 5~GHz, Xiang et al. 2006), which is probably because of
the large Faraday depth in the line of sight, implying that their
host AGN are largely edge-on to us.

Complete samples of GPS sources are statistically vital for
studying the early evolution of radio-loud AGN and the triggering
of their nuclear activity, such as the 1 Jy complete GPS sample in
the northern sky (Stanghellini et al. 1998), the Parkes
half-Jansky (PHJ) sample in the southern/equatorial sky (Snellen
et al. 2002), and the CORALZ sample at low redshift (Snellen et
al. 2004). The Very Long Baseline Interferometry (VLBI) technique
is a key tool for studying the morphologies of GPS sources from
the GPS samples. We defined a subsample of twelve GPS sources with
declination $>-5^{\circ}$ that had not been observed with the VLBI
before, from the PHJ sample, and carried out the VLBI observations
at 1.6 GHz and 5 GHz of them in 2006 and 2008, respectively. It is
found that most of the sources exhibit compact double structure by
analysing the 1.6 GHz European VLBI Network (EVN) observation (Liu
et al. 2007, hereafter paper I). In the follow-up 5 GHz EVN
observation, we demonstrate that we can resolve their VLBI
structure more sharply and obtain the spectral indices of
components between 1.6 GHz and 5 GHz, and finally identify CSOs
from the candidates.

Quite a few GPS sources having been found to have flux variability
(Torniainen et al. 2007). Since 2007, we have performed flux
monitoring at 4.85 GHz with the Urumqi 25-m radio telescope for
GPS sources from the master list for which declination
$>-30^{\circ}$ (Labiano et al 2007).

Throughout the paper, the cosmological parameters $H_{0}=70.1$ km
s$^{-1}$ Mpc$^{-1}$, $\Omega_{\Lambda}=0.721$ and
$\Omega_{m}=0.279$ (Komatsu et al. 2009) are adopted, and the
spectral index $\alpha$ is defined by $S\propto\nu^{-\alpha}$.

\section{The EVN observations and data reduction}

The 5 GHz VLBI observations were carried out on 12 June 2008 for
the first nine targets in Table~\ref{table:gps} (proposal code
EL036A) and on 25 October 2008 for the remaining three (EL036B),
using the MK5 recording system with a bandwidth of 32 MHz and a
sample rate of 256 Mbps in dual circular polarization. The EVN
antennae involved were Effelsberg, Westerbork, Jodrell (Jb1),
Medicina, Noto, Onsala, Torun, Hartebeesthoek, Urumqi, and
Shanghai. Hartebeesthoek was not involved in EL036B because of an
antenna problem. The snapshot observations of the target sources
(see Table~\ref{table:gps}) were carried out in a total of about
24 hours, DA193 and OQ208 were inset as calibrators. The data
correlation was performed at the Joint Institute of VLBI in Europe
(JIVE).

The Astronomical Image Processing System (AIPS) software has been
used for editing, apriori calibration, fringe-fitting,
self-calibration, imaging, and model-fitting of the data. The
uncertainty in the flux density in the VLBI image is estimated to
be about 10\%, and the total time per source and noise level in
source image are listed in Table~\ref{table:vlbi}.

\section{Results and comments on individual sources}

We list basic information about 12 GPS sources and some parameters
derived from the VLBI images in Table~\ref{table:gps}. Columns 1
through to 13 provide source names, optical identification (G:
galaxy, QSO: quasar), redshift (de Vries et al. 2007 and the
literature, those with * are a photometric estimated by Tinti et
al. 2005), linear scale factor pc/mas where $H_{0}=100h$ km
s$^{-1}$ Mpc$^{-1}$, maximum angular size from the 5 GHz VLBI
observation, maximum linear size, 4.85 GHz total flux density
measured with Urumqi single dish in December 2008 (July 2007 value
for J0210+0419), the resolved-out fraction in the 5 GHz VLBI
image, the VLBI component, the integrated flux density of the
component at 1.6 GHz (from paper I), the integrated flux density
of the component at 5 GHz (using the same restoring beam as that
used in the 1.6 GHz image), and the spectral index of component
calculated from Columns (11) and (12). More image parameters are
listed in Table~\ref{table:vlbi}.

In the following, we summarize the VLBI results and comment on the
individual sources.

\begin{table*}

         \caption[]{The GPS sample.}
         $$
\begin{tabular}{cccccccccccccc}
            \hline
            \hline
            \noalign{\smallskip}
            1&2&3&4&5&6&7&8&9&10&11&12&13\\
IAU&Other&Id&z&Scale &$\theta_{5GHz}$&$LS$&$S_{4.85Ur}$&$Resolved$&$Comp.$&$S_{int}^{1.65}$&$S_{int}^{5}$&$\alpha_{1.65}^{5}$\\
name  & &   &   &(pc/mas)& (mas)  & (pc) & (mJy)   &      (\%)  &       &  (mJy)         & (mJy)       &      \\
            \noalign{\smallskip}
            \hline
            \noalign{\smallskip}

J0210+0419   &B0208+040& G   & 1.5*  & 6.11$h^{-1}$    & 84     & 513$h^{-1}$  & 282$\pm$2    & $\sim$14    & A     & 375  & 144 &0.86  \\
            &   &        &       &        &        &      &                  &          & B       & 205     & 99    & 0.66 \\
J0323+0534  &4C+05.14 & G   &  0.1785& 2.13$h^{-1}$    &48      & 102$h^{-1}$  & 796$\pm$21      & $\sim$83    & A  & 1270  & 102 &2.27  \\
&&&&&&&&                                                                                   & B      & 548     & --     &  --       \\
J0433$-$0229 &4C$-$02.17& G   &  0.53 & 4.47$h^{-1}$    & 72     & 322$h^{-1}$  & 612$\pm$5      & $\sim$27    & A  & 1045  & 396 &0.88  \\
&&&&&&&&                                                                                   & B    & 111     & 16  &  1.75 \\
J0913+1454  &B0910+151& G   &  0.47* & 4.19$h^{-1}$    & 76     & 318$h^{-1}$  & 275$\pm$18        & $\sim$8    & A  & 501 & 208 & 0.79  \\
&&&&&&&&                                                                                  & B      & 157   & 45   & 1.13   \\
J1057+0012  &B1054+004 & G   & 0.65* & 4.97$h^{-1}$    & 13     & 65$h^{-1}$   & 355$\pm$21      & $\sim$11  & A  & 550  & 299  & 0.55  \\
&&&&&&&&                                                                                   & B    & 58   & --     & --   \\
J1109+1043  &B1107+109& G   &  0.55* & 4.62$h^{-1}$    & 63    & 290$h^{-1}$  & 420$\pm$7       & $\sim$8     & A  & 984 & 281  & 1.13  \\
&&&&&&&&                                                                                   & B    & 311  & 91   & 1.11  \\
J1135$-$0021 &4C$-$00.45 & G   & 0.975 & 5.75$h^{-1}$    & 110    & 633$h^{-1}$  & 422$\pm$13       & $\sim$15  & A  & 454  & 196 & 0.76  \\
&&&&&&&&                                                                                    & B    & 271  & 123  & 0.71 \\
J1203+0414   &B1200+045& QSO  &  1.221 & 5.96$h^{-1}$    & 72     & 429$h^{-1}$  & 590$\pm$21       & $\sim$22   & A & 850  & 423 & 0.63  \\
&&&&&&&&                                                                                   & B     & 81   & 24    & 1.10    \\
&&&&&&&&                                                                                   & C    & 35   & 14    & 0.83   \\
J1352+0232  &B1349+027 & G   & 0.607 & 4.83$h^{-1}$    & 188    & 908$h^{-1}$ & 458$\pm$10        & $\sim$56   & A  & 480  & 189   &0.84  \\
&&&&&&&&                                                                                   & B    & 120  & 13    &2.00  \\
J1352+1107  &4C+11.46 & G   & 0.891 & 5.61$h^{-1}$    & 21     & 118$h^{-1}$  & 403$\pm$ 6       & $\sim$75    &all & 598 & 102 & 1.59  \\
J1600$-$0037 &B1557$-$004& G   &       &             & 65     &      & 190$\pm$5        & $\sim$11    & A    & 607 & 85   & 1.77  \\
&&&&&&&&                                                                                  & B     & 255 & 75   & 1.10   \\
&&&&&&&&                                                                                  & C     & --   & 4    &   --     \\
J2058+0540 &4C+05.78 & G   &  1.381 & 6.04$h^{-1}$    & 153    & 923$h^{-1}$ & 351$\pm$18        & $\sim$37    & A  & 513 & 134 & 1.21  \\
&&&&&&&&                                                                                  & B     & 403  & 86   & 1.39  \\
            \noalign{\smallskip}
            \hline
        \end{tabular}{}
         \label{table:gps}
         $$
   \end{table*}

   \begin{table*}
         \caption[]{The component parameters of the VLBI images at 5 GHz. The columns provide: (1)
         source name and classification (CSOc: CSO candidate, cj: core-jet, cx:
         complex); (2) total observing time per source in minutes; (3) noise level per beam in 5 GHz VLBI
         image;
         (4) total VLBI flux density; (5) component identification labelled to paper I; (6), (7) peak and
         integral intensity of a fitted Gaussian component at 5 GHz with the AIPS task JMFIT; (8), (9) major/minor
         axes and position angle of the component at 5 GHz; (10), (11) distance and position angle relative
         to the first component; and (12) brightness temperature of the component.}
         $$
\begin{tabular}{ccccccccccccc}
          \hline
           \hline
            \noalign{\smallskip}

1     &2      & 3       &4        &  5       &   6                    & 7      & 8       & 9           & 10  & 11 & 12     \\
$Name$   &$T$&$RMS$&$S_{vlbi}$ & $Comp.$   &  $S_{p}$  &  $S_{int}$  & $\theta_{1}\times\theta_{2}$ & $PA$      & $D$       & $PA$  & $T_{b}$  \\
\& Class &(min.)& (mJy/b) & (mJy)&          &   (mJy/b)   &   (mJy)       &     (mas$\times$mas)           & ($^\circ$)  & (mas)    & ($^\circ$) &($\times10^{8}K$)\\

            \noalign{\smallskip}
            \hline
            \noalign{\smallskip}

J0210+0419 &50&0.8&243 &A       &94      &141     & 3.2 $\times$ 2.0             &1.4       &0              &                   &11.8         \\
CSO        &   &  &   &B1      &28      &65      & 4.2 $\times$ 3.8             &176      &71.4 $\pm$ 0.1 &-152.7 $\pm$ 0.1    &2.2          \\
           &   &  &  &B2      &22      &63      & 6.3 $\times$ 2.9             &38       &67.1 $\pm$ 0.1 &-152.7 $\pm$ 0.1    &1.8  \\
J0323+0534 &149&0.2&135 &A1      &19      &19       & 5.0 $\times$ 1.0             &83       &0              &                   &0.3      \\
CSOc       &   &   &  &A2      &42      &109     & 16.8 $\times$ 5.1            &59        &5.2 $\pm$ 0.1  &42.9 $\pm$ 0.1    &0.3  \\
           &   &   &  &A3      &8       &16       & 12.9 $\times$ 3.8            &63        &22.8 $\pm$ 0.1 &57.4 $\pm$ 0.1    & 0.1       \\
J0433$-$0229&162&0.8& 449 &A1      &270     &388      & 5.0 $\times$ 3.8    &161       &0          &             &6.7          \\
 CSOc/cj    &   &   &    &A2      &30      &35       & 3.7 $\times$ 1.3             &157       &13.0 $\pm$ 0.1 &166.4 $\pm$ 0.1   &2.4   \\
            &   &   &   &B       &9       &15       & 7.9 $\times$ 3.2             &156      &52.6 $\pm$ 0.1 &154.8 $\pm$ 0.1    &0.2     \\
J0913+1454 &108&0.2& 253  &A1      &24      &59       & 2.9 $\times$ 1.5            &105       &0              &                   &4.0    \\
  CSOc     &&  & &A2      &11      &37       & 3.7 $\times$ 2.6             &116       &1.4 $\pm$ 0.1  &-81.8 $\pm$ 0.4   &1.0        \\
           &&  & &A3      &3       &5        & 3.5 $\times$ 1.6             &169       &5.5 $\pm$ 0.1  &-77.0 $\pm$ 0.4   &0.3       \\
           &&  & &A4      &17      &41       & 2.7 $\times$ 2.3             &72        &10.6 $\pm$ 0.1 &-85.5 $\pm$ 0.1   &2.0        \\
           &&  & &A5      &16      &41       & 3.0 $\times$ 2.0             &94        &11.5 $\pm$ 0.1 &-86.9 $\pm$ 0.1   &2.0      \\
           &&  & &A6      &11      &21       & 2.2 $\times$ 1.6             &69         &5.6 $\pm$ 0.1  &91.6 $\pm$ 0.1   &2.0   \\
           &&  & &A7      &3       &9        & 4.2 $\times$ 3.0             &25         &6.6 $\pm$ 0.1  &91.7 $\pm$ 0.4   &0.2       \\
           &&  & &A8      &2       &3        & 2.8 $\times$ 1.1             &37        &11.8 $\pm$ 0.1 &88.8 $\pm$ 0.3    &0.3     \\
           &&  & &B       &19      &47       & 3.0 $\times$ 0.6             &117        &54.1 $\pm$ 0.1 &71.7 $\pm$ 0.1   &8.0       \\
J1057+0012 &84&0.3& 316  &A       &183     &289      & 3.1 $\times$ 1.4         &36        &0          &            & 23.0    \\
    cj     &  &  & &        &        &         &                              &           &              &                  &            \\
J1109+1043 &90&0.2& 386  &A       &173     &278      & 4.4 $\times$ 1.7             &58        &0           &                     & 12.0    \\
  CSOc     &  &   & &B       &55      &88       & 4.3 $\times$ 1.7             &65        &46.3 $\pm$ 0.1 &103.3 $\pm$ 0.1   &4.0     \\
J1135$-$0021&60&0.7&358  &A1      &134     &185      & 3.3 $\times$ 2.2             &171       &0            &                    &11.0       \\
   CSO      && & &A2      &29      &69       & 7.4 $\times$ 2.4             &135       &3.3 $\pm$ 0.5  &165.0 $\pm$ 4.2   &1.6   \\
            && & &A3      &9       &27       & 10.5 $\times$ 2.2            &41        &14.5 $\pm$ 1.9 &165.5 $\pm$ 3.5   &0.5    \\
            && & &B1      &44      &79       & 6.0 $\times$ 2.4             &176       &94.6 $\pm$ 0.3 &163.3 $\pm$ 0.1   &2.3        \\
            && & &B2      &35      &91       & 8.8 $\times$ 3.0             &158       &91.4 $\pm$ 0.5 &162.7 $\pm$ 0.1   &1.5    \\
            && & &B3      &19      &69       & 13.6 $\times$ 2.6            &154       &86.6 $\pm$ 1.2 &163.6 $\pm$ 0.2   &0.8   \\
J1203+0414 &48&0.2& 461  &A1      &260     &396      & 2.5 $\times$ 1.1             &118       &0       &                         & 68.4  \\
  cj       &&  & &A2      &25      &56       & 4.5 $\times$ 1.3             &126       &4.3 $\pm$ 0.1  &119.5 $\pm$ 0.1   &2.0   \\
           &&  & &B       &11      &15       & 1.9 $\times$ 1.1             &94        &14.6 $\pm$ 0.1 &101.4 $\pm$ 0.1   &3.0  \\
           &&  & &C       &8       &12       & 2.8 $\times$ 0.4             &111       &63.3 $\pm$ 0.1 &104.4 $\pm$ 0.1   &5.0    \\
J1352+0232 &45&0.2& 202  &A1      &90      &106      & 3.6 $\times$ 1.7             &95        &0      &      &    6.0    \\
  CSOc     &&&    &A2      &57      &68       & 4.3 $\times$ 1.7             &59         &7.1 $\pm$ 0.1 &-133.6 $\pm$ 0.1   &3.0    \\
           &&  & &B       &10      &13       & 4.9 $\times$ 1.2             &168       &170.1 $\pm$ 0.1 &-112.6 $\pm$ 0.1  & 0.8  \\
J1352+1107 &54&0.5& 102  &A       &29      &77       & 9.2 $\times$ 3.7             &138       &0       &        &1.0 \\
  cx       && &  &B       &13      &60       & 14.3 $\times$ 5.4            &39        &5.2 $\pm$ 0.1 &56.4 $\pm$ 0.1      & 0.3    \\
           && &  &C       &15      &46       & 12.4 $\times$ 2.8            &20        &4.8 $\pm$ 0.1 &-39.7 $\pm$ 0.1     &0.5   \\
J1600$-$0037&103&0.2& 169 &A       &70      &83       & 2.5 $\times$ 2.2             &130       &0      &                           &   \\
 cj         &&&  &B1      &29      &57       & 10.0 $\times$ 1.0            &69         &31.8 $\pm$ 0.1 &84.6 $\pm$ 0.1    &    \\
            && & &B2      &21      &46       & 10.9 $\times$ 1.2            &80        &25.2 $\pm$ 0.1 &87.5 $\pm$ 0.1     &    \\
            && & &C       &4       &4        & 1.8 $\times$ 1.1             &72         &50.8 $\pm$ 0.1 &77.8 $\pm$ 0.1    &  \\
J2058+0540 &47&0.4& 220  &A1      &104     &124      & 4.5 $\times$ 1.0             &9.3       &0       &                          & 14.0   \\
   CSO     &&  & &A2      &8.2     &13       & 7.7 $\times$ 2.1             &177       &15.5 $\pm$ 0.1 &-175.0 $\pm$ 0.2   & 0.4     \\
           &&  & &B       &42      &68       & 8.3 $\times$ 2.4             &172       &132.6 $\pm$ 0.1&171.9 $\pm$ 0.1    & 2.0   \\

\noalign{\smallskip}
            \hline
            \end{tabular}{}
             \label{table:vlbi}
              $$
\end{table*}

\subsection{J0210+0419}

\begin{figure}
     \centering
     \includegraphics[width=7.5cm]{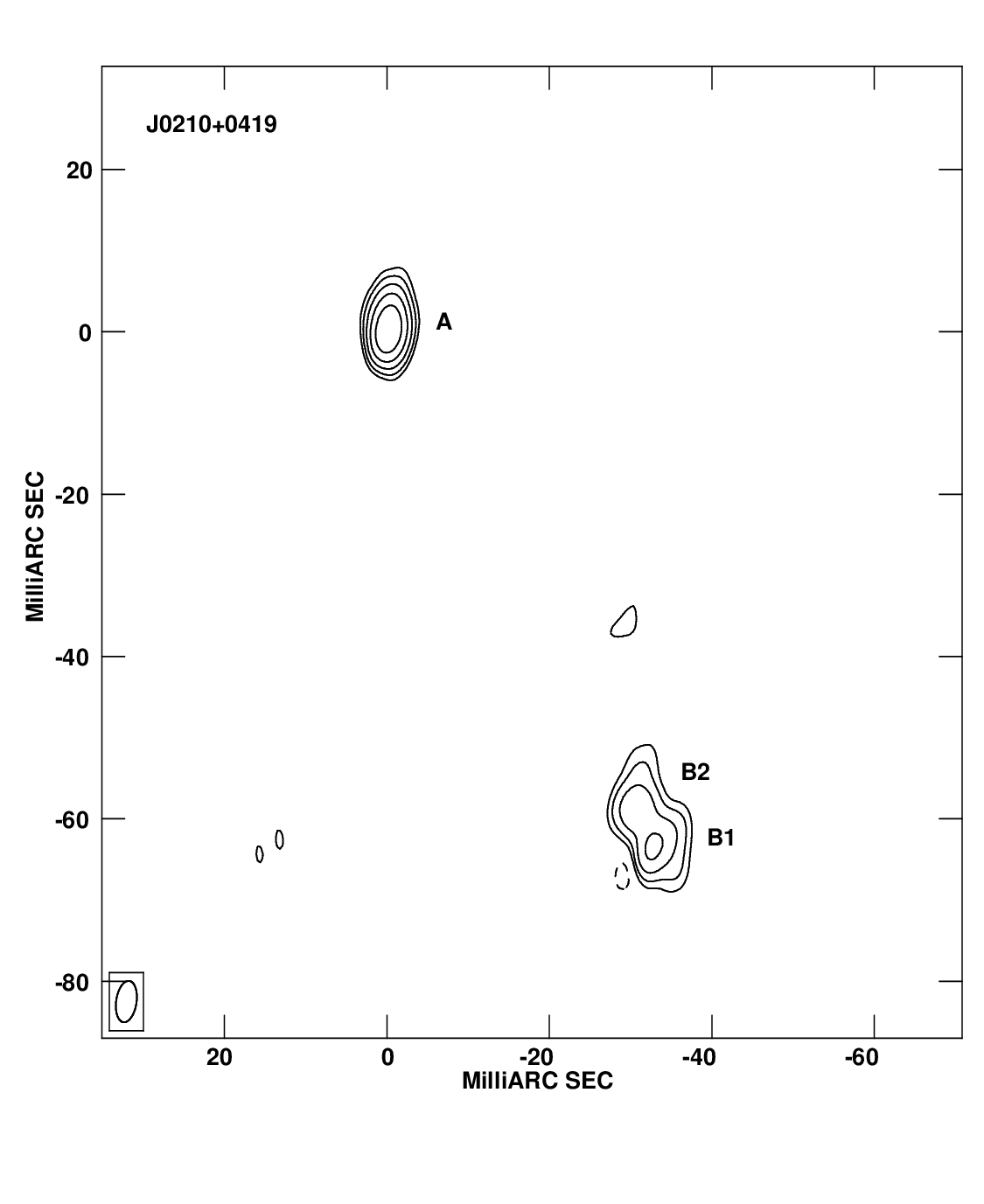}
     \caption{J0210+0419: the restoring beam is
        $5.1\times2.5$ mas in PA $-8.2^{\circ}$, the contours
        are 3 mJy/beam times -1, 1, 2, 4, 8, and 16, and the peak flux density is 95 mJy/beam.
       }
      \label{fig1}
   \end{figure}

The 5 GHz VLBI image (Fig.~\ref{fig1}) of this source shows a
double lobe structure, which is similar to the 1.6 GHz image in
paper I. The north-east lobe is more compact at 5 GHz, and the
south-west one is resolved into `head-tail' components `B1' and
`B2'. The spectral indices of component `A' and `B' between 1.6
GHz and 5 GHz are both steeper (0.86 and 0.66, respectively, see
Table~\ref{table:gps}), indicating that they are lobes/hotspots.
The core of the source is not detected. The total flux density is
slightly lower than for PKS90 data at 5 GHz in
Table~\ref{table:flux}. According to the double lobes and tail
morphologies at both 1.6 GHz and 5 GHz and the steep spectra of
the lobes, we classify the source as a CSO.

\subsection{J0323+0534}

\begin{figure}
     \centering
     \includegraphics[width=7.5cm]{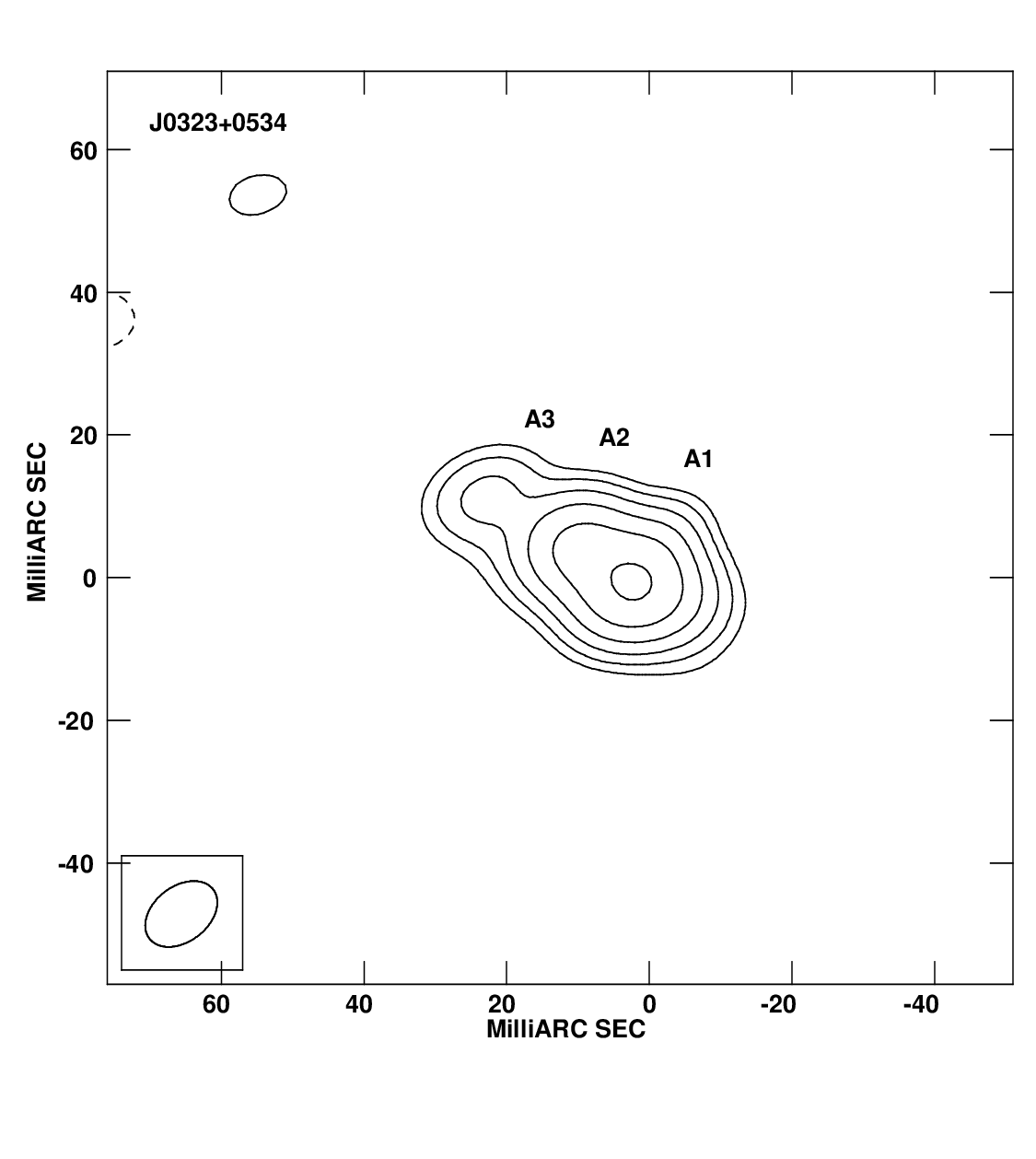}
     \caption{J0323+0534: the restoring beam is
        $11.3\times7.7$ mas in PA $-52.3^{\circ}$, the contours
        are 1.5 mJy/beam times -1, 1, 2, 4, 8, 16, and 32, and the peak flux density is 54 mJy/beam.}
      \label{fig2}
   \end{figure}

The 83\% of the total flux density at 5 GHz
(Table~\ref{table:gps}) is resolved-out in the 5 GHz VLBI image
(Fig.~\ref{fig2}), confirming that the source is very diffuse as
suggested in paper I. The southern diffuse component `B' in the
1.6 GHz VLBI image (paper I) is resolved-out completely in the 5
GHz VLBI image. The north region `A', which is very diffuse at 1.6
GHz, is resolved into three components `A1', `A2', and `A3', which
appear to have a `core-jet' shape at 5 GHz; this leads to a very
steep spectral index $\alpha_{1.65}^{5}$ of the component `A', but
`A1', `A2, and `A3' are most likely types of hotspots rather than
a core-jet. The total flux density is stable compared to PKS90
data at 5 GHz in Table~\ref{table:flux}, hence we keep this source
as a CSO candidate.

\subsection{J0433$-$0229}

\begin{figure}
     \centering
     \includegraphics[width=7.5cm]{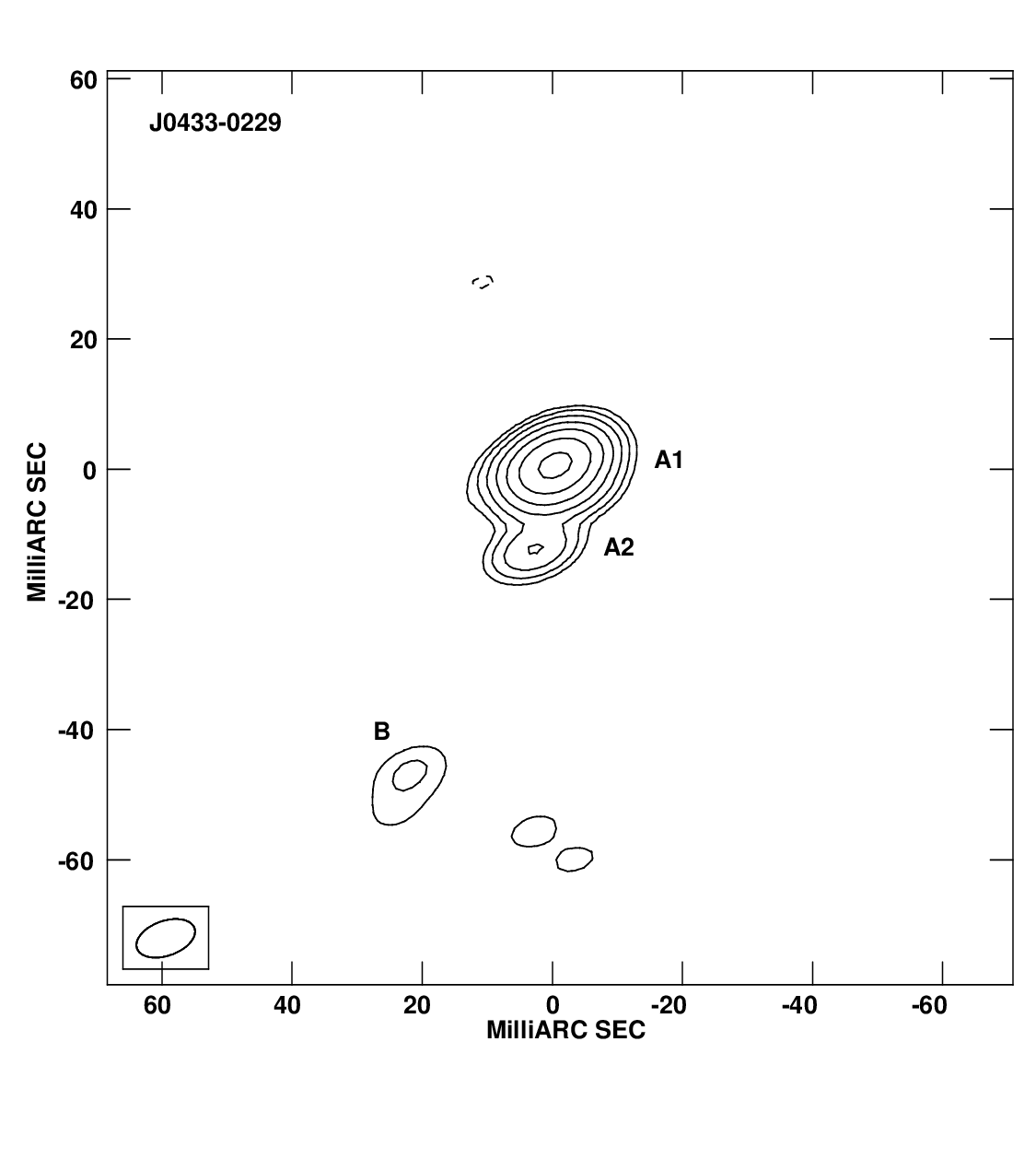}
     \caption{J0433$-$0229: the restoring beam is
        $9.4\times5.3$ mas in PA $-70.5^{\circ}$, the contours
        are 3.5 mJy/beam times -2, 1, 2, 4, 8, 16, 32, and 64, and the peak flux density is 270 mJy/beam.}
      \label{fig3}
   \end{figure}

The major component `A' of 1.6 GHz image (paper I) is resolved
into `A1' and `A2' at 5 GHz (Fig.~\ref{fig3}), and the weak
component `B' is marginally detected. The overall shape of the
source at 5 GHz is similar to that at 1.6 GHz, resembling  a
core-jet. Both of the components `A' and `B' however, have steep
spectra (Table~\ref{table:gps}), and about 27\% of the total flux
density is resolved-out in the 5 GHz image. The total flux density
is stable relative to PKS90 data at 5 GHz in Table 3. This source
may be a core-jet one or a CSO candidate, although additional high
sensitivity VLBI observation would be needed to clarify this.

\subsection{J0913+1454}

\begin{figure}
     \centering
     \includegraphics[width=7.5cm]{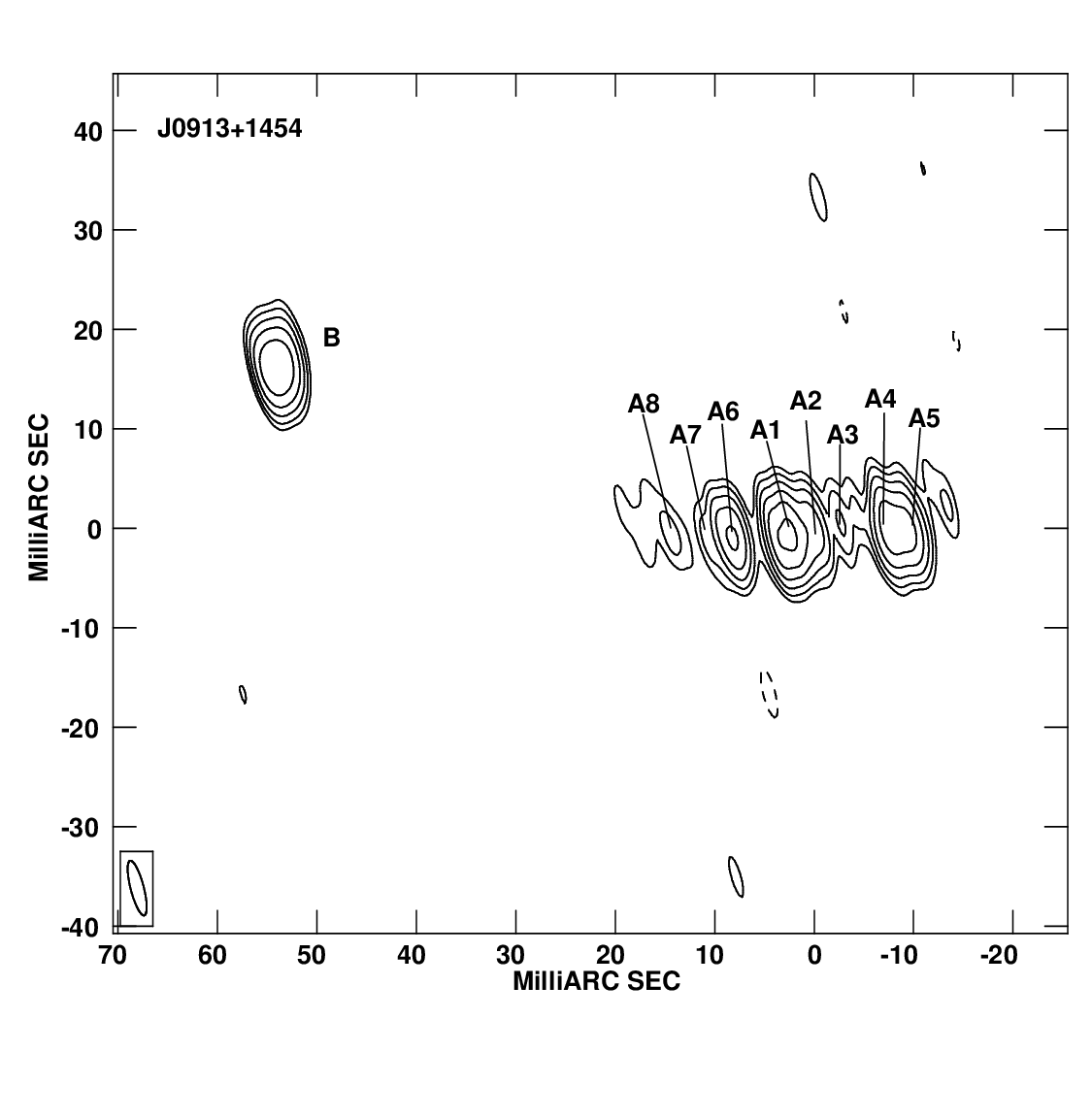}
     \caption{J0913+1454: the restoring beam is
        $5.7\times1.3$ mas in PA $14.5^{\circ}$, the contours
        are 0.6 mJy/beam times -2, 1, 2, 4, 8, 16, and 32, and the peak flux density is 24 mJy/beam.}
      \label{fig4}
   \end{figure}

The 5 GHz VLBI image (Fig.~\ref{fig4}) of this source can be
resolved into several components `A1' to `A8' in the region `A',
and the component `B', which is more compact at 5 GHz than at 1.6
GHz (paper I). It is difficult to classify the source, the
components `A1' to `A8' most likely representing a core and a set
of jets. The core may be at either `A8' or `A5': in the former
case the source may be a CSO, in the latter the source may be a
core-jet. The spectral index $\alpha_{1.65}^{5}$ of component `A'
(integrated) and `B' is 0.79 and 1.13, respectively. The total
flux density is stable compared to PKS90 data at 5 GHz in
Table~\ref{table:flux}.

\subsection{J1057+0012}

\begin{figure}
     \centering
     \includegraphics[width=7.5cm]{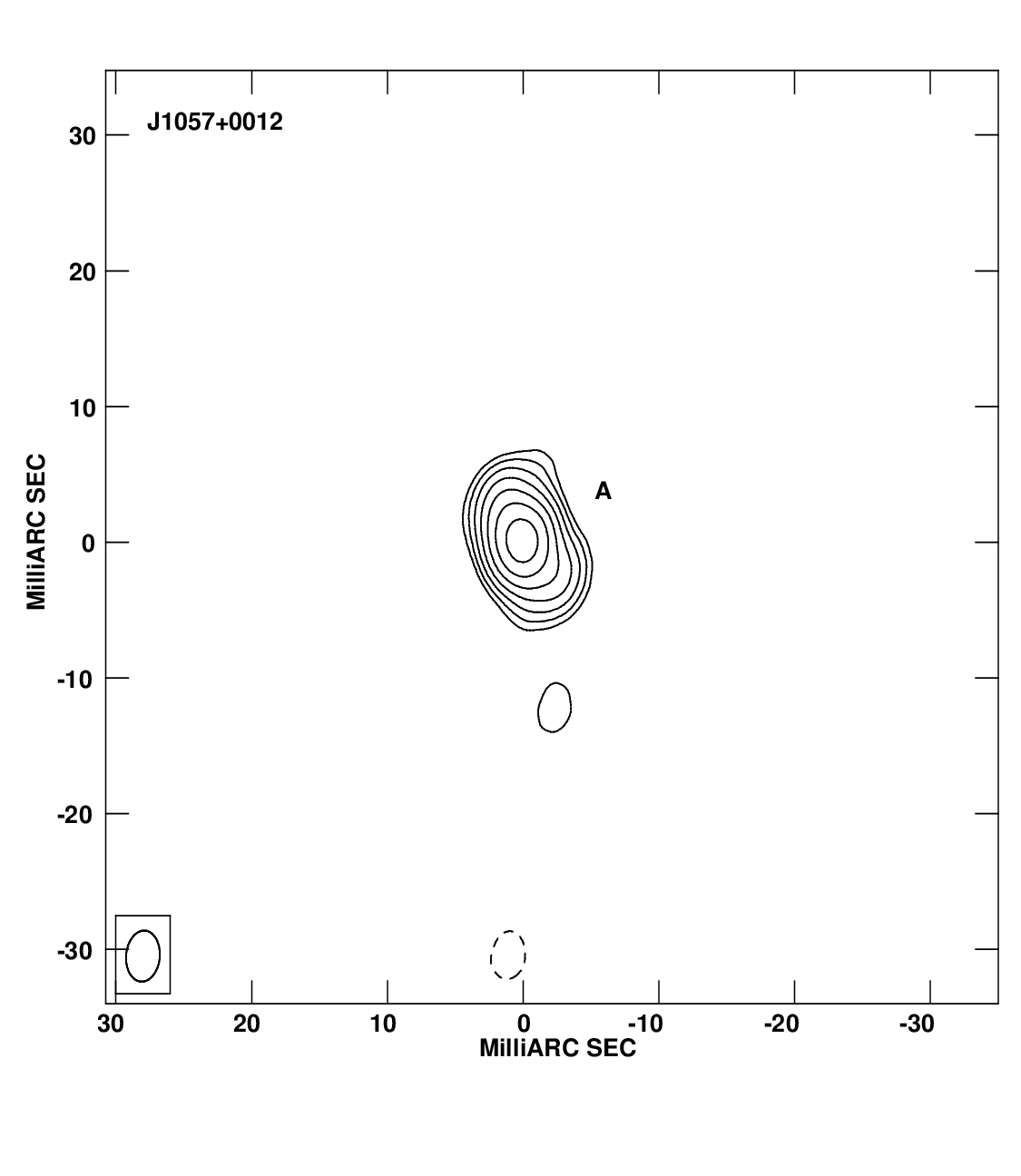}
     \caption{J1057+0012: the restoring beam is
        $3.8\times2.5$ mas in PA $-4.7^{\circ}$, the contours
        are 2 mJy/beam times -1, 1, 2, 4, 8, 16, 32, and 64, and the peak flux density is 192 mJy/beam.}
      \label{fig5}
   \end{figure}

The 5 GHz image (Fig.~\ref{fig5}) contains a point-like source,
the possible jet in the 1.6 GHz image (paper I) being neither
detected nor resolved-out. The spectral index $\alpha_{1.65}^{5}$
of main component is 0.55, which is flatter than that of a typical
lobe component in other sources. The total flux density is
slightly lower than that of the PKS90 data at 5 GHz in
Table~\ref{table:flux}. We classify it as a core-jet source on the
basis of both the 1.6 GHz and 5 GHz images.

\subsection{J1109+1043}

\begin{figure}
     \centering
     \includegraphics[width=7.5cm]{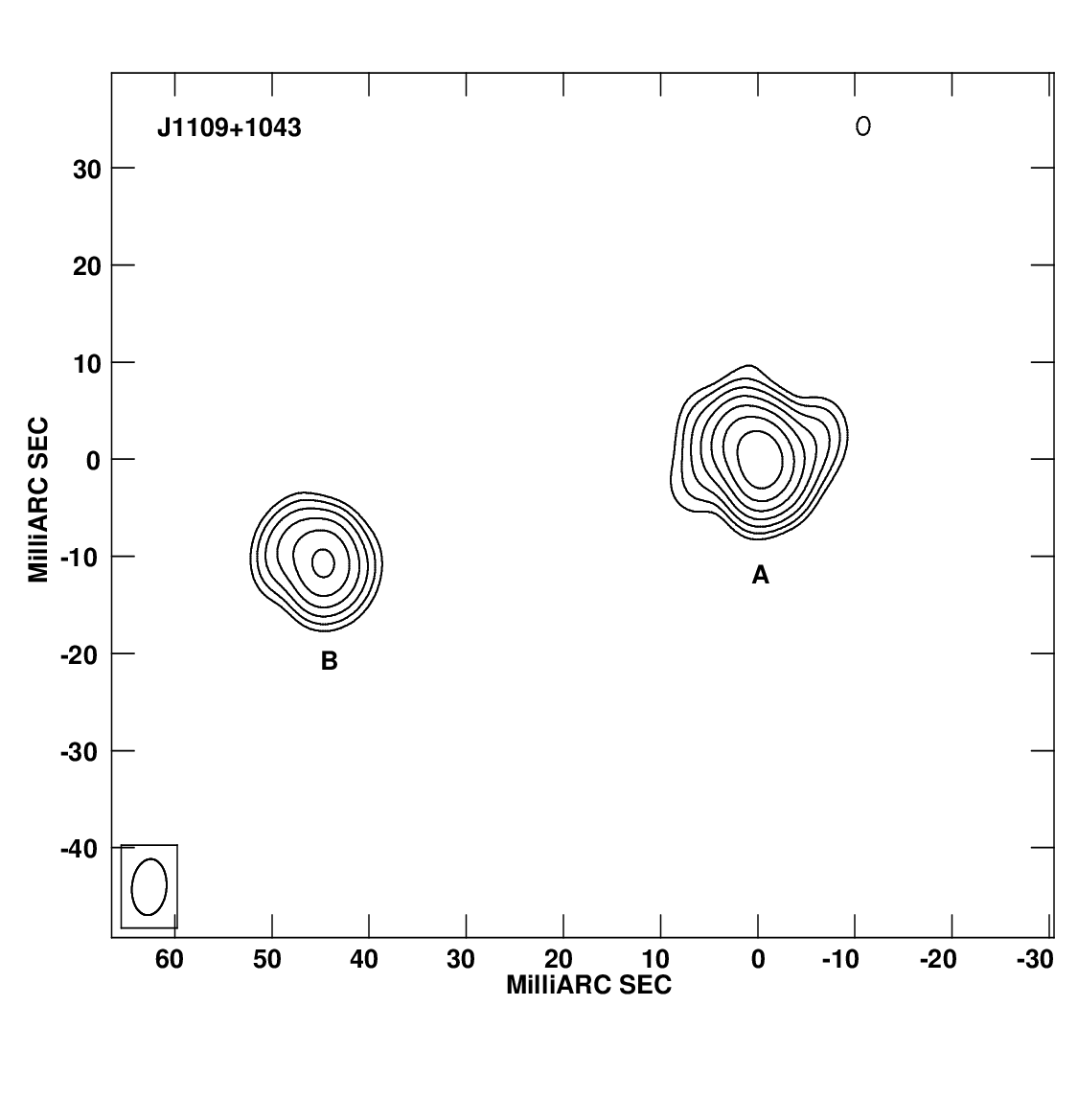}
     \caption{J1109+1043: the restoring beam is
        $5.8\times3.6$ mas in PA $-5^{\circ}$, the contours
        are 1.5 mJy/beam times -1, 1, 2, 4, 8, 16, 32, and 64, and the peak flux density is 178 mJy/beam.}
      \label{fig6}
   \end{figure}

The 5 GHz VLBI image (Fig.~\ref{fig6}) of this source exhibits a
double structure, which is similar to the 1.6 GHz image (paper I).
Both components have equally steep spectra between 1.6 GHz and 5
GHz, of index 1.13 for component `A' and 1.11 for component `B',
whose values indicate that they are lobs/hotspots. The total flux
density is slightly higher than that of the PKS90 data at 5 GHz in
Table~\ref{table:flux}. The source is most likely a CSO or a
compact double (CD) since no tail/jet emission has been detected.

\subsection{J1135$-$0021}

The 5 GHz VLBI image (Fig.~\ref{fig7}) shows a double-lobe
structure, which is similar to that in 1.6 GHz image (paper I).
The lobes/tails can be fitted by several subcomponents (`A1',
`A2', `A3' for `A', and `B1', `B2', `B3' for `B'). The spectral
indices of `A' and `B' between 1.6 GHz and 5 GHz are equally
steep, and the linear size is smaller than 1 kpc $h^{-1}$
(Table~\ref{table:gps}). The total flux density is slightly lower
than for the PKS90 data at 5 GHz in Table~\ref{table:flux}. We
confirm that the source is a CSO.

\subsection{J1203+0414}

This is a GPS quasar, and the 5 GHz VLBI image (Fig.~\ref{fig8})
exhibits a core-jet like structure. The component `A1' probably
presents the core of the source, since the spectral index 0.63 of
`A' corresponds to a flatter spectrum than those of other
components `B' and `C'. The total flux density is higher by 15.3\%
than the PKS90 data at 5 GHz in Table~\ref{table:flux}, and it is
resolved-out by about 22\% in the 5 GHz VLBI image, some weak jets
possibly being resolved. We classify the quasar as a core-jet
source.

\subsection{J1352+0232}

As in the 1.6 GHz VLBI image in paper I, the 5 GHz VLBI image
(Fig.~\ref{fig9}) exhibits a strong component `A' and a weak
component `B'. The bright component `A' can be fitted with two
close components `A1' and `A2'. The components `A' and `B' have
the integrated spectral indices $\alpha_{1.65}^{5}$ of 0.84 and
2.0, respectively. The total flux density of the source is
resolved-out by about 56\% in the 5 GHz VLBI image. The total flux
density is stable compared with PKS90 data at 5 GHz in
Table~\ref{table:flux}. The source could be a CSO if it were
really a double-lobe source, or a core-jet source if the core
resided in the component `A1'.

\subsection{J1352+1107}

The 5 GHz VLBI image of the source detected a complex feature
(Fig.~\ref{fig10}), which is fitted with three components `A',
`B', and `C'. About 75\% of the total flux density
(Table~\ref{table:gps}) is resolved-out in the 5 GHz VLBI image,
confirming that the source is very diffuse as suggested in paper
I. The whole source exhibits a steep spectrum of 1.59 between 1.6
GHz and 5 GHz. The total flux density is stable compared to PKS90
data at 5 GHz in Table~\ref{table:flux}. With the current data in
hand, it is difficult to classify this source; both a core and
curving jets are probably embedded in the diffuse emission.

\subsection{J1600$-$0037}

The 5 GHz VLBI image (Fig.~\ref{fig11}) has a core-jet like
feature, which consists of a bright component `A', components
`B1'/`B2', and a weak component `C'. The spectral index of the
component `A' between 1.6 GHz and 5 GHz is 1.77. It is difficult
to identify the core because the `B' is not clearly resolved in
the 1.6 GHz VLBI image (paper I), the component `B1' in the 5 GHz
VLBI image probably being a candidate core. The weak component `C'
emerging in the 5 GHz image, which is not detected in 1.6 GHz
image (probably because of low-frequency absorption), is possibly
a counter-jet. The total flux density has a variability of 7.5\%
in 2007--2009 and a variation of 7.2\% relative to PKS90 data at 5
GHz in Table~\ref{table:flux}. We therefore classify the source as
a core-jet source on the basis of its VLBI morphology and flux
variability.

\subsection{J2058+0540}

The 5 GHz VLBI image (Fig.~\ref{fig12}) shows a double-lobe
source, which is similar to the 1.6 GHz image in paper I. The
northern lobe is resolved into `head-tail' components `A1' and
`A2' in the 5 GHz image. The spectral indices of components `A'
and `B' between 1.6 GHz and 5 GHz are 1.21 and 1.39, respectively.
The total flux density is stable relative to the PKS90 data at 5
GHz in Table~\ref{table:flux}. We classify the source as a CSO. We
found that 37\% of the total flux density is resolved-out in the 5
GHz image, indicating that in the source some form of jet and/or
diffuse emission is missing in the VLBI image.

 \begin{table*}
         \caption[]{Flux density and possible variability of the 12 GPS sources.
         Column 2 is the PKS90 flux density at 5.0 GHz. Columns 3 to 7
         indicate the
         4.85 GHz flux density measured with the Urumqi telescope in January 2007,
         July 2007, December 2008, July 2009, and November 2009, respectively; Column 8-10
         give the mean flux density of the Urumqi data, the standard deviation of the Urumqi data
         divided by the mean value, and the relative variation between the mean flux density
         and the PKS90 data.
         }
         $$
         \begin{tabular}{cccccccccc}

            \hline
            \noalign{\smallskip}
$Source$ & $S_{PKS}$ & $S_{Jan07}$ & $S_{Jul07}$ & $S_{Dec08}$ & $ S_{Jul09}$ & $S_{Nov09}$&$\overline{S_{Ur}}$ & $\sigma_{S_{Ur}}/\overline{S_{Ur}}$ & $(\overline{S_{Ur}}-S_{PKS})/S_{PKS}$\\

&(mJy)&(mJy)& (mJy) & (mJy) & (mJy) & (mJy)& (mJy) &(\%) & (\%) \\
            \noalign{\smallskip}
            \hline
            \noalign{\smallskip}

J0210+0419 &  300  &302$\pm10$&$282\pm2$& &$274\pm4$& $298\pm4$&289&4.0 &$-$3.7 \\

J0323+0534 &  830&868$\pm$9& $833\pm5$&$796\pm21$&$814\pm10$& $836\pm4$ &829&2.9&$-$0.1\\

J0433$-$0229 & 640 &637$\pm$14&$705\pm25$&$612\pm5$&&  &651&6.0&1.8\\

J0913+1454 & 300 &297$\pm$8&$289\pm3$&$275\pm18$&& $311\pm6$ &293&4.5&$-$2.3\\

J1057+0012& 370 & 351$\pm$6 & $369\pm4$ & $355\pm21$ & $345\pm4$ & $373\pm14$ &359&3.0&$-$3.1\\

J1109+1043& 400 & 428$\pm$8 & 416$\pm$5 & 420$\pm$7 & 434$\pm$4 & $414\pm7$ &422&1.8&5.6\\

J1135$-$0021 &440  &427$\pm$8 &$392\pm4$ & $422\pm13$  &  &   &414&3.7 &$-$6.0 \\

J1203+0414&  520& 611$\pm$7 &$561\pm3$& $590\pm21$ &$604\pm4$& $632\pm5$ &600&3.9&\textbf{15.3} \\

J1352+0232 & 470 &  $469\pm7$  &$453\pm4$   & $458\pm10$ &$470\pm4$& $465\pm6$ &463&1.4&$-$1.5 \\

J1352+1107 & 410& 418$\pm$5 &$406\pm3$ & $403\pm6$ &$411\pm3$& $432\pm3$&414&2.5&1.0 \\

J1600$-$0037 & 180   & 212$\pm$3&$177\pm5$& $190\pm5$  & &   &193&\textbf{7.5}&\textbf{7.2}\\

J2058+0540 & 340 &  &$340\pm3$ & $351\pm18$   &  $343\pm4$ & $345\pm5$ &345&1.2&1.4\\

           \noalign{\smallskip}
            \hline
           \end{tabular}{}
         $$
         \label{table:flux}
   \end{table*}

\section{Flux density observations of the sources at 4.85 GHz}

At high radio frequencies in particular, it has been reported
that, many GPS/HFP sources exhibit significant flux variability
(Torniainen et al. 2007, Orienti et al. 2007). Jauncey et al.
(2003) found that $\sim10\%$ of GPS sources experienced flux
density variability over a period of 30 months. We measured the
flux densities of the GPS sources in the Labiano et al. (2007)
sample at 4.85 GHz, by observing with the Urumqi 25-m telescope
between 2007 and 2009. Compared to 87GB data, Liu et al. (2009)
found that 44\% GPS quasars exhibit $>10\%$ flux variation over a
period of about 20 years, while this fraction is only 11\% for GPS
galaxies. Detailed analysis of our monitoring data will be
presented in another paper, and here we just list the measured
flux densities of our VLBI targets in Table~\ref{table:flux} for
epochs during which most of the VLBI targets are observed.

The flux densities were determined using antenna slews with `cross
- scans' in azimuth and elevation, fourfold in each coordinate.
This enabled us to check the pointing offsets in both coordinates.
After applying a correction for small pointing offsets, the
amplitudes of both AZ and EL were averaged. We then corrected the
measurements for the elevation-dependent antenna gain and the
remaining systematic time-dependent effects, using a number of
steep spectrum and non-variable secondary calibrators. Finally, we
related our observations to the absolute flux density by using the
scale 7.5 Jy at 5 GHz of the primary calibrator 3C286 (Ott et al.
1994).

Given the flux variation measured in 2007--2009, and by comparing
with PKS90 5 GHz data in Table~\ref{table:flux}, we find that the
flux densities are quite stable or slightly different for the
majority of the twelve GPS sources, but that two core-jet sources
J1203+0414 (quasar) and J1600$-$0037 exhibit considerable
variability in total flux density.

\section{Discussion}

It is interesting to estimate the jet-viewing angles for the
confirmed CSOs, because the CSOs are assumed to lie nearly within
the plane of sky. From the VLBI morphologies and the steep
spectral indices of the VLBI components between 1.6 GHz and 5 GHz,
we can firmly classify J0210+0419, J1135$-$0021, and J2058+0540 as
CSOs that display symmetric double lobes/tails. We estimated the
jet viewing angle from the flux ratio between the approaching and
receding lobes with the formula (Taylor \& Vermeulen 1997)
\begin{displaymath}
\frac{S_{a}}{S_{r}}=\left(\frac{1+\beta\cos\Theta}{1-\beta\cos\Theta}\right)^{k+\alpha},
\end{displaymath}
where $S_{a}/S_{r}$ is the flux ratio between the approaching
(stronger) lobe and receding lobe, $\Theta$ is the viewing angle
(degree) of the jet axis, $\alpha$ is the source spectral index
$\alpha_{h}$ in optical-thin regime (from paper I), $k=2$ or $k=3$
for a continuous or discrete jet, respectively, and $\beta$ is the
lobe velocity as a fraction of the speed of light. By adopting
$\beta=0.1$, which is a mean lobe velocity in CSOs (Giroletti \&
Polatidis 2009), we estimated the jet viewing angles for the
continuous and discrete cases at 1.6 GHz and 5 GHz, respectively,
in Table~\ref{table:angle}. From a rough estimation, we could say
that the three CSOs have relatively large jet-viewing angles.
Although this method should be used only for the confirmed
double-lobe sources, and the lobes of the CSOs are not resolved
too much in the VLBI images.

Nevertheless the majority of sources in our sample have
double-lobe like morphologies and total flux densities that are
quite stable over 20 years. Only 3 CSOs are firmly classified
according to Fanti's suggestion that a CSO should have a core
in-between double lobes or have double lobes with twin jets/tails
if a core is not detected (Fanti 2009).

Furthermore, 6 sources are resolved-out in the 5 GHz VLBI images
by more than 20\% of total flux density, probably due to diffuse
emission in the sources. This may lead to quite steep spectral
indices of VLBI components of the sources as listed in
Table~\ref{table:gps}. The spectra of all 6 sources peak at 0.4
GHz (Snellen et al. 2002), which is close to the spectral peaks of
CSS sources (see e.g., Fanti et al. 1995), so some of the sources
are probably not GPS but rather CSS sources if they have
larger-scale diffuse emission that is resolved-out by the VLBI
observations.

\begin{table}
\caption{The jet viewing angles estimated for the CSOs.}
\label{table:angle}
\begin{tabular}{lccc|cc}
\hline
\noalign{\smallskip}
Source  & Type  & $\Theta_{1.6G}^{k=2}$ & $\Theta_{1.6G}^{k=3}$  & $\Theta_{5G}^{k=2}$  & $\Theta_{5G}^{k=3}$   \\
\noalign{\smallskip}
\hline
\noalign{\smallskip}
J0210+0419   & CSO & -- & 37  &  48  & 60 \\
J1135$-$0021 & CSO & 22 & 47  &  33  & 52 \\
J2058+0540   & CSO & 66 & 72  & 41   & 56 \\
\noalign{\smallskip} \hline
\end{tabular}
\end{table}

\section{Summary and conclusion}

We have presented the results of 5 GHz EVN observations of 12 GPS
sources from the sample of Snellen (2002) for the first time. The
source structure and spectral indices between 1.6 GHz and 5 GHz
are obtained. We have also carried out total flux density
measurements with the Urumqi 25-m telescope at 4.85 GHz in
2007--2009, and compared the mean flux density with the PKS90
data.

\begin{enumerate}

\item From the source morphologies, the component spectral indices
and the total flux variability, we can firmly classify three
sources J0210+0419, J1135$-$0021, and J2058+0540 as CSOs, and
J1057+0012, J1203+0414, and J1600$-$0037 as core-jet sources. The
others J0323+0534, J0433$-$0229, J0913+1454, J1109+1043, and
J1352+0232 remain as CSO candidates, and J1352+1107 is a complex
VLBI feature whose nature need to clarify in the future.

\item The total flux densities of the GPS sources at 5 GHz were quite stable
in 2007--2009 and over the long term compared with the PKS90 data,
except for core-jet sources.

\item We estimated the jet viewing angles $\Theta$ for the CSOs
by using the double-lobe flux ratio. The three CSOs were found to
have systematically larger $\Theta$.

\item We plan to observe the CSOs in the future with the VLBI at 5 GHz
to measure the expansion speeds of lobes and estimate the ages of these
young radio sources.

\end{enumerate}

\begin{acknowledgements}

We thank the anonymous referee for valuable comment, and Prof.
D.-R. Jiang and Prof. X.-W. Cao for comments on the manuscript.
The European VLBI Network is a joint facility of European,
Chinese, South African and other radio astronomy institutes funded
by their national research councils. This work is supported by the
National Natural Science Foundation of China (NNSFC) under grant
No.10773019 and the National Basic Research Program of China (973
Program 2009CB824800).

\end{acknowledgements}

\clearpage

\clearpage

\begin{figure}
     \centering
     \includegraphics[width=7.5cm]{14075fg7.eps}
     \caption{J1135$-$0021: the restoring beam is
        $5.5\times3.5$ mas in PA $5.7^{\circ}$, the contours
        are 2.8 mJy/beam times -1, 1, 2, 4, 8, 16, and 32, and the peak flux density is 137 mJy/beam.}
      \label{fig7}
   \end{figure}

\begin{figure}
     \centering
     \includegraphics[angle=-90,width=7.5cm]{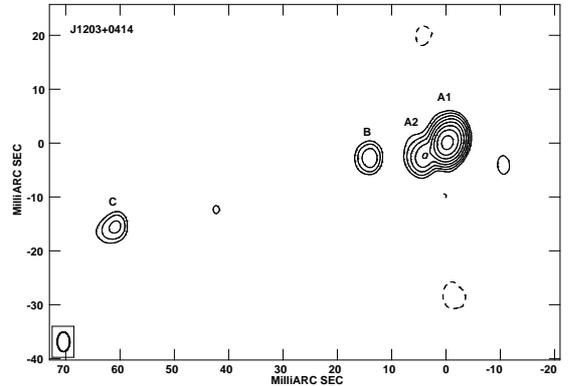}
     \caption{J1203+0414: the restoring beam is
        $3.7\times2.3$ mas in PA $0.9^{\circ}$, the contours
        are 1.5 mJy/beam times -1, 1, 2, 4, 8, 16, 32, 64, and 128, and the peak flux density is 262 mJy/beam.}
      \label{fig8}
   \end{figure}

\clearpage

\begin{figure}
     \centering
     \includegraphics[angle=-90,width=7.5cm]{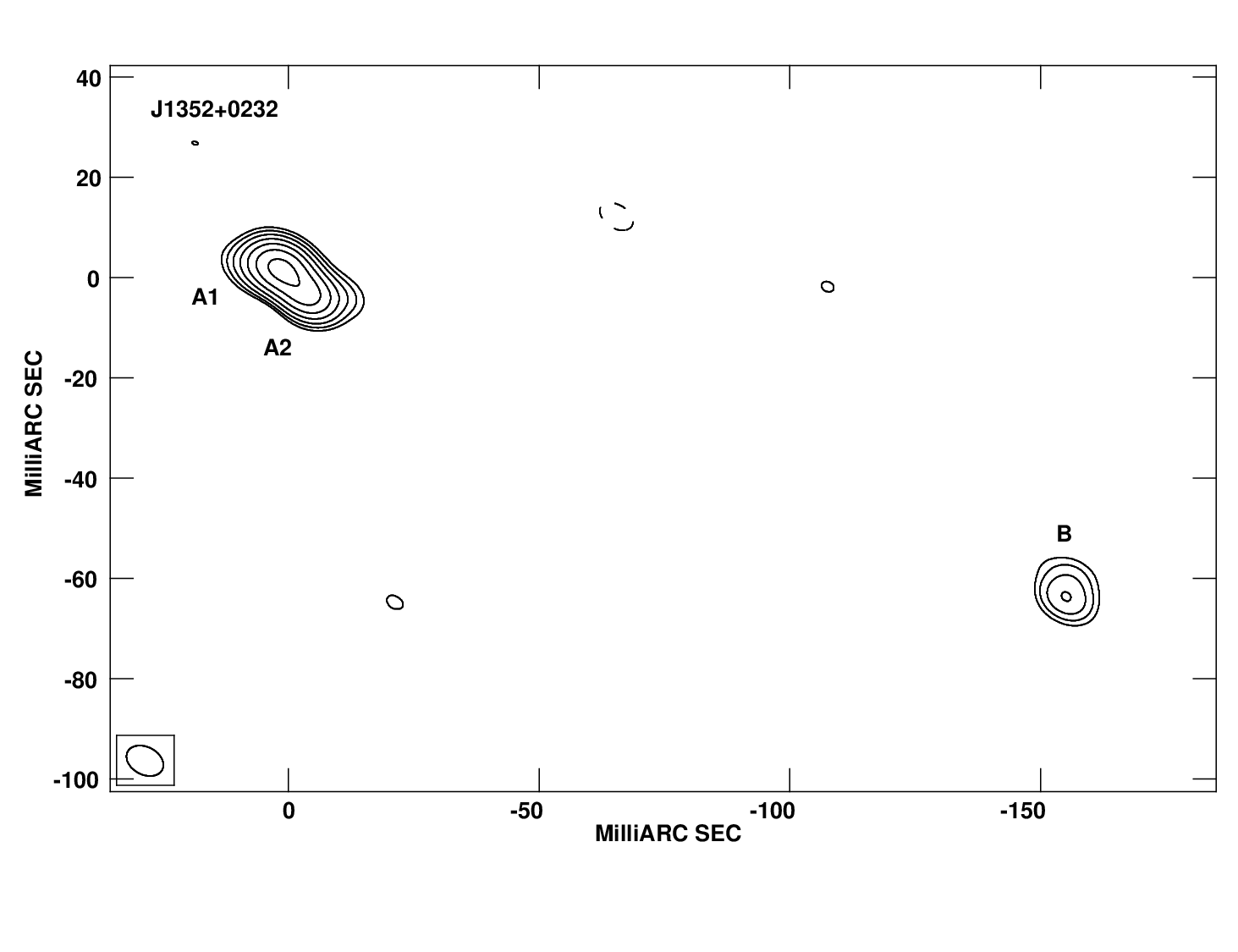}
     \caption{J1352+0232: the restoring beam is
        $7.8\times5.5$ mas in PA $62.3^{\circ}$, the contours
        are 1.2 mJy/beam times -1, 1, 2, 4, 8, 16, 32, and 64, and the peak flux density is 102 mJy/beam.}
      \label{fig9}
   \end{figure}

\begin{figure}
     \centering
     \includegraphics[width=7.5cm]{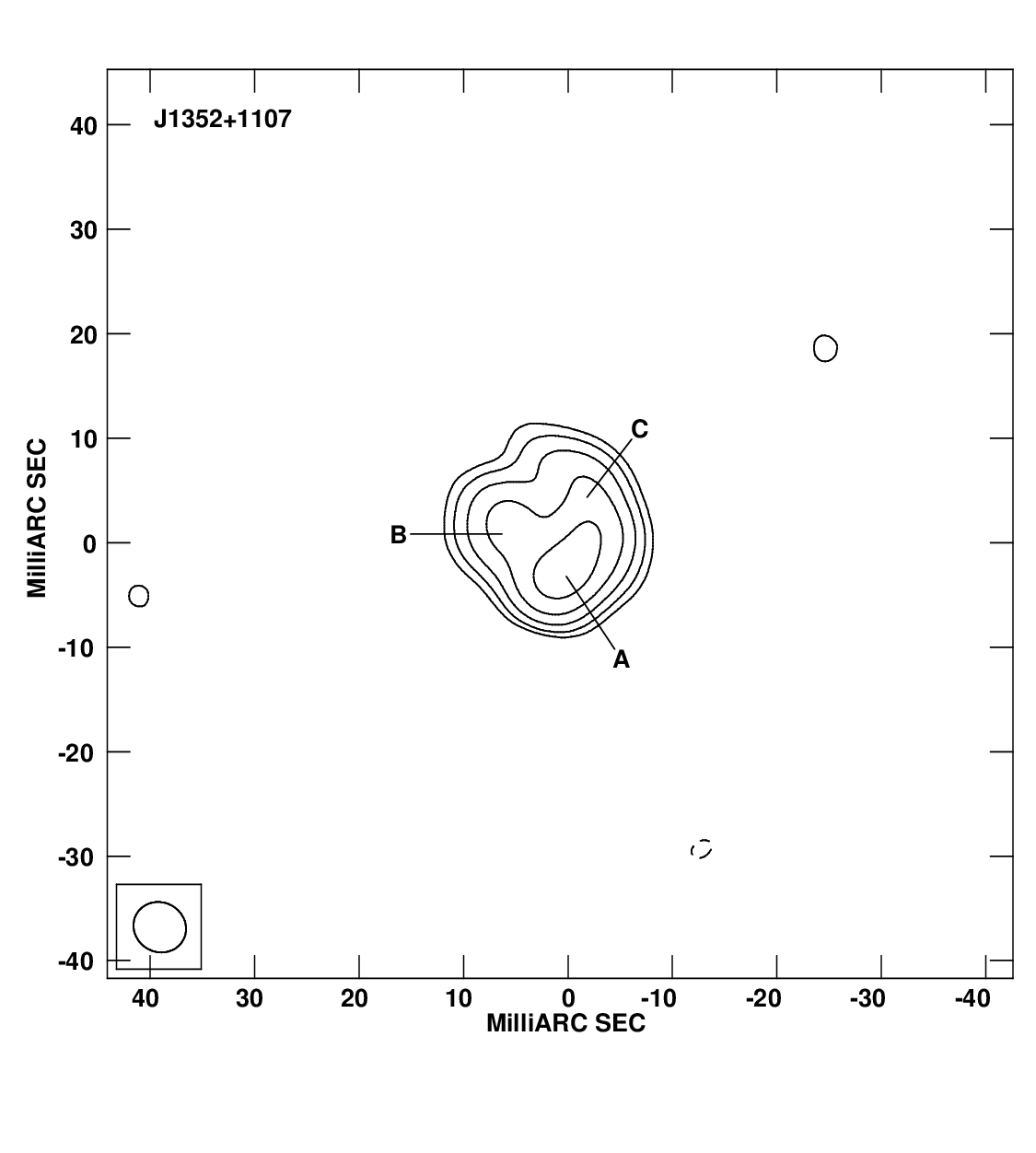}
     \caption{J1352+1107: the restoring beam is
        $5.1\times4.7$ mas in PA $61.1^{\circ}$, the contours
        are 1.2 mJy/beam times -1, 1, 2, 4, 8, and 16, and the peak flux density is 29 mJy/beam.}
      \label{fig10}
   \end{figure}

\begin{figure}
     \centering
     \includegraphics[width=7.5cm]{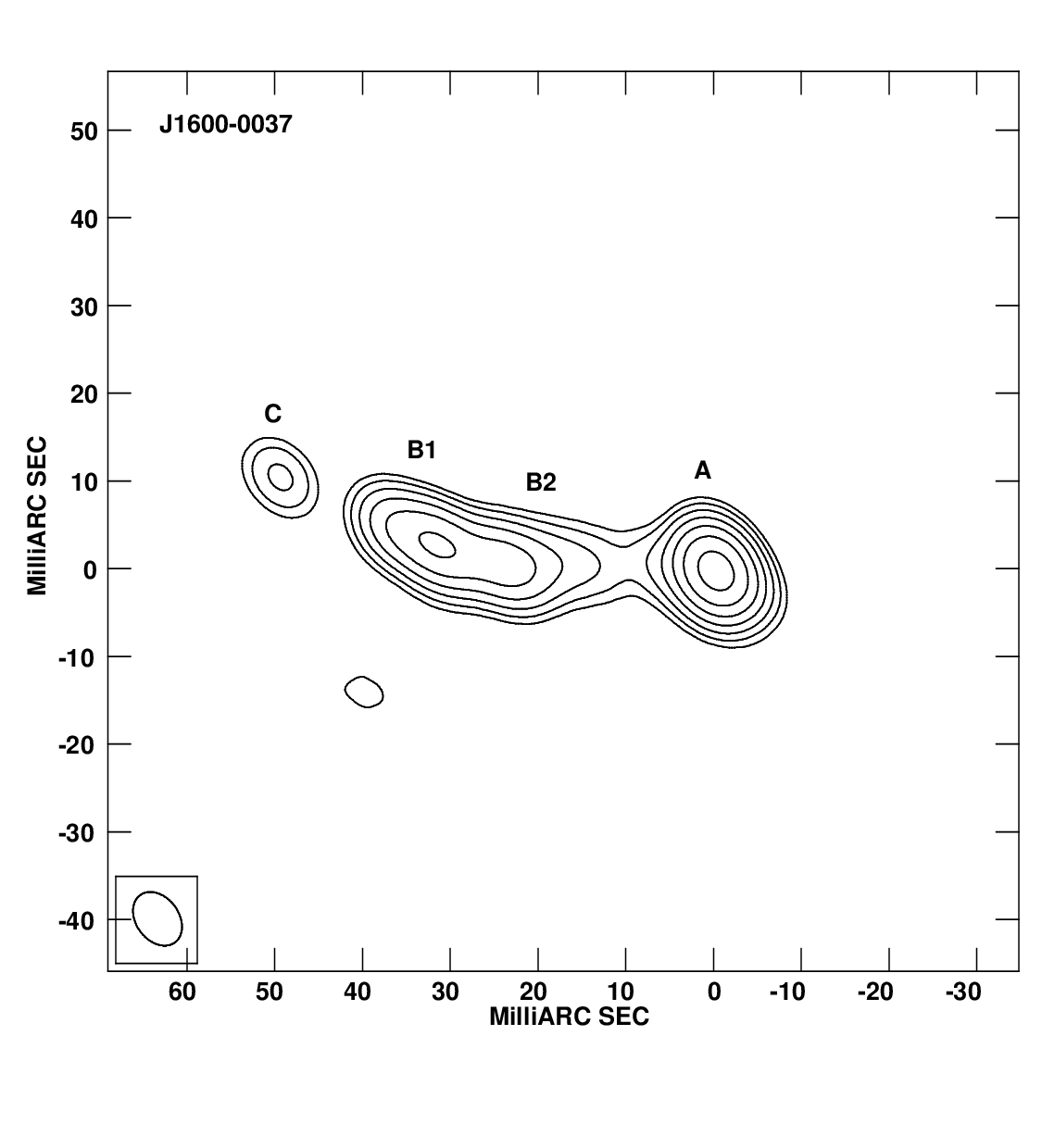}
     \caption{J1600$-$0037: the restoring beam is
        $6.7\times4.9$ mas in PA $36.3^{\circ}$, the contours
        are 0.8 mJy/beam times -1, 1, 2, 4, 8, 16, 32, and 64, and the peak flux density is 71 mJy/beam.}
      \label{fig11}
   \end{figure}

\begin{figure}
     \centering
     \includegraphics[width=7.5cm]{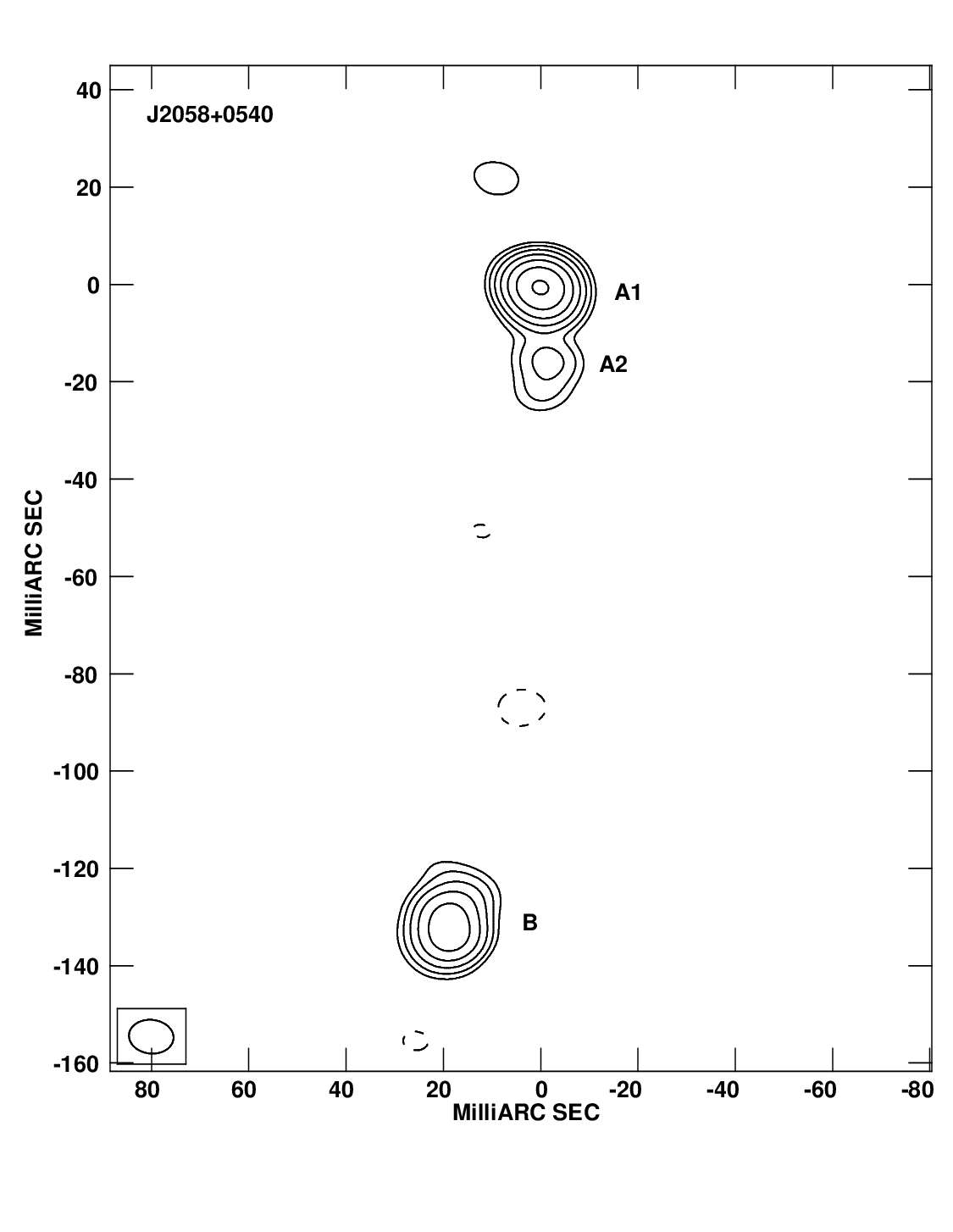}
     \caption{J2058+0540: the restoring beam is
        $9.2\times6.9$ mas in PA $84.2^{\circ}$, the contours
        are 1.5 mJy/beam times -1, 1, 2, 4, 8, 16, 32, and 64, and the peak flux density is 105 mJy/beam.}
      \label{fig12}
   \end{figure}

\end{document}